\documentclass[twocolumn,nofootinbib,
showpacs,aps,floatfix]{revtex4}
\usepackage{bm}
\usepackage{amssymb}
\usepackage{graphicx}

\def\beq{\begin{equation}}
\def\eeq{\end{equation}}
\def\beqa{\begin{eqnarray}}
\def\eeqa{\end{eqnarray}}
\newcommand {\fexp} [1] {\exp \left( #1 \right)}
\newcommand {\fabsq}[1] {\left| #1 \right|^2}
\newcommand {\fabs}[1] {\left| #1 \right|}

\newcommand {\si} {\, \mbox{s}^{-1}}
\newcommand {\mum}{\, \mu \mbox{m}}
\newcommand {\cms}{\, \mbox{cm/s}}

\begin{document}
\title{Three-dimensional effects in ``atom diodes'':
atom-optical devices for one-way motion}

\author{A. Ruschhaupt}
\email[Email address: ]{a.ruschhaupt@tu-bs.de}
\affiliation{Institut f\"ur Mathematische Physik, TU Braunschweig, Mendelssohnstrasse 3, 38106 Braunschweig, Germany}

\author{J. G. Muga}
\email[Email address: ]{jg.muga@ehu.es}
\affiliation{Departamento de Qu\'\i mica-F\'\i sica, Universidad del
Pa\'\i s Vasco, Apdo. 644, 48080 Bilbao, Spain}

\begin{abstract}
The ``atom diode'' is
a laser device that lets the ground state
atom pass in one direction but not 
in the opposite direction. We examine three-dimensional
effects of that device    
for arbitrary atomic incidence angles on flat laser sheets
and set breakdown limiting angles and velocities. It is found that a correct
diodic behavior independent of
the incident angle can be obtained with blue 
detuned lasers.
\end{abstract}
\maketitle

\section{Introduction}

\begin{figure}[t]
\begin{center}
\includegraphics[angle=0,width=0.8\linewidth]{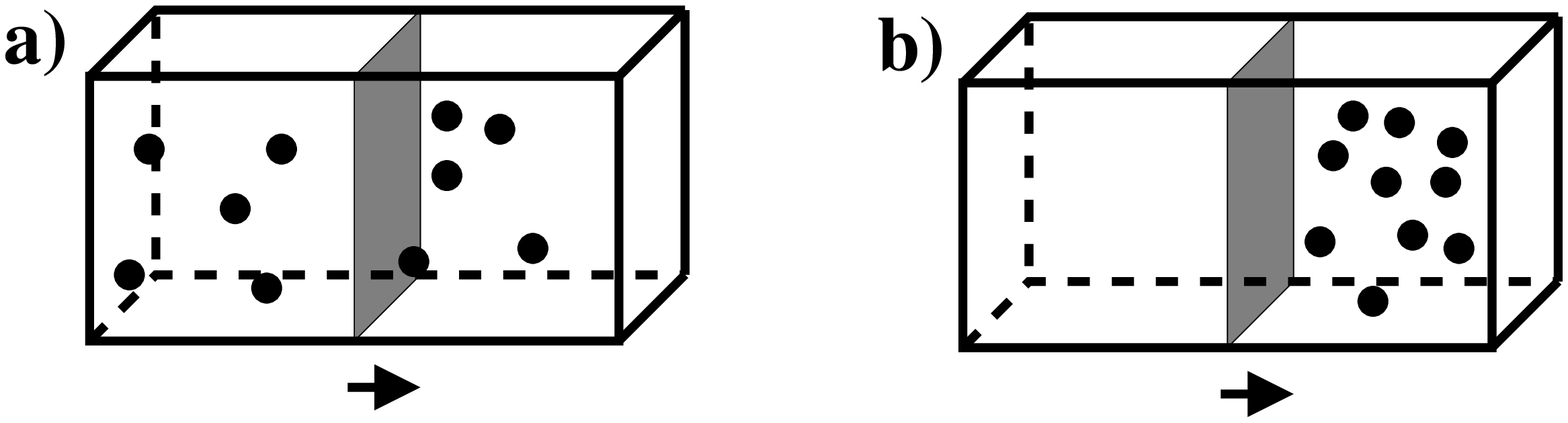}
\end{center}
\caption{\label{fig1}A version of Maxwell's demon:
the atom diode would achieve a differential of pressure 
as an inanimate one-way valve, by letting atoms cross in one direction but
not in the opposite one.}
\end{figure}

Atom optics is ``basically the
development and perfection of techniques and devices which can manipulate the atom
waves coherently'' \cite{rev}. Many of these devices (lenses, mirrors, beam splitters) have their origin 
in standard light optics, and others are influenced by electronics: this has lead to the idea of guiding the atoms in atomic circuits and developing 
circuit control elements.  
We have proposed in particular, in analogy with the electronic diode, an
``atom diode''
\cite{ruschhaupt_2004_diode,ruschhaupt_2006_diode,ruschhaupt_2006_quench},   
a laser device which combines state-selective mirrors and pumping
lasers to let the neutral atom in its ground state pass in one direction 
(conventionally from left to right), but not 
in the opposite direction for a range of incident velocities.  
Similar ideas have been put forward independently by Raizen and
coworkers \cite{raizen.2005,dudarev.2005} who proposed one-way barriers
to implement a new laser cooling technique sweeping them 
across an atomic trap. The near possibility of an experimental implementation 
has motivated a recent analysis on the effect of quenching lasers
to improve the diode performance \cite{ruschhaupt_2006_quench}, as well as  
the present work, which investigates three dimensional (3D) effects that 
had been ignored so far, and the way to overcome them.     
The original models have been one-dimensional, which is appropriate for tight 
lateral confinement in wave guides or cigar shaped traps, but for general 
geometries, and in particular for flat laser sheets and arbitrary incidence 
angles, three dimensional effects may become important as we shall see.
As a further motivation, consider       
an interesting application of the atom diode shown schematically
in Fig. \ref{fig1}, namely a realization of Maxwell's pressure demon
\cite{maxwell,leff1,leff2,ruschhaupt_2006_demon}:
Starting with a box of atoms at random positions we put an atom
diode in the middle (the direction of transmission  is
indicated by an arrow). After some time, all atoms will be on
the right-hand side of the box provided that 
the diode works independently of atomic velocity and incident angle. 
%
%

The paper begins with an analysis of the $2$-level diode in Sec. II, and
Sec. III discusses the $3$-level diode: this is a worthwhile complication since 
it provides a simple physical realization 
of the former case with just two lasers by adiabatic elimination.       

Throughout the paper, flat laser sheets are assumed to extend  
in the $y-z$-plane with some thickness in $x$-direction,
and laser wave-vectors point in $y$-direction. 
Because of translational symmetry in $z$-direction and
the corresponding conservation of $z$-momentum, this coordinate 
will be ignored hereafter.

\section{Atom diode based on $2$ levels plus quenching}

%
\begin{figure}[t]
\begin{center}
\includegraphics[angle=0,width=0.8\linewidth]{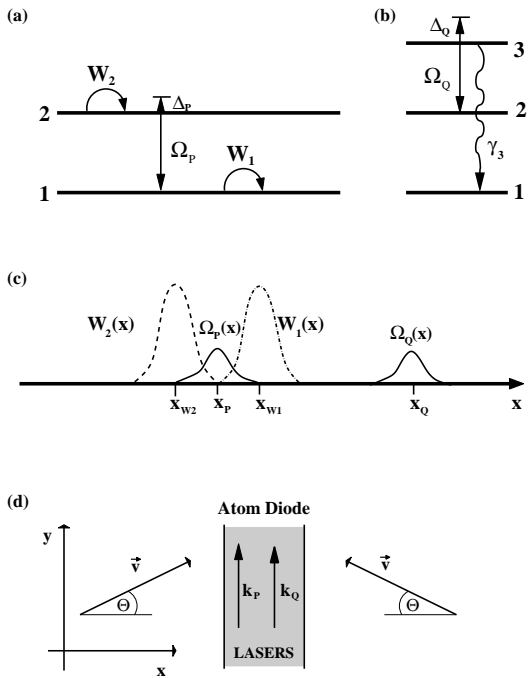}
\end{center}
\caption{\label{fig2}Atom diode based on $2$ levels plus quenching:
(a,b) Schematic action of the different lasers 
on the atom levels, 
(c) location of the different laser potentials in $x$ direction, and
(d) setting in the $x-y$ plane.}
\end{figure}
%

The basic setting for a two-level atom diode can be seen
in Fig. \ref{fig2}a and \ref{fig2}c, and consists of 
three, partially overlapping, interaction regions: in the external ones, 
state-selective mirrors block the 
excited ($|2\rangle$) and ground  ($|1\rangle$)
state atoms, respectively;  
in the central pumping region, a laser, possibly detuned with respect to the atomic transition
with detuning $\Delta_P$ (the difference between the pumping laser frequency and
atomic transition frequency $\omega_{21}$ between levels 1 and 2) couples the two levels. 
The pumping laser has a wave vector of modulus $k_P>0$ in $y$ direction and the Rabi
frequency is $\Omega_P$. The detuning, which is a novel feature
with respect to our previous work, plays an important role in 3D as we shall see.  
In principle it is possible to implement this scheme 
with three different lasers, one for each region, nevertheless a 
simpler realization relies on just two lasers 
appropriately detuned and a STIRAP (Stimulated Raman Adiabatic Passage
\cite{bergmann.1998}) process, this case    
is worked out in Sec. III as a generalization of the results
discussed in \cite{ruschhaupt_2006_diode}.
  
The explanation of the diodic behavior is the following (see Fig. \ref{fig2}c):
If the ground-state atom travels from the right to left and its velocity is
not too high, it is reflected by the state-selective mirror
potential $W_1 \hbar/2$ and returns to the right.
If the atom travels from the left in the ground state,
it will cross the potential $W_2 \hbar/2$, be pumped to the second level
and pass to the right unaffected by the potential $W_1 \hbar/2$.
It might seem that $W_2\hbar/2$ plays no role but, if it is removed, the
transmittance $1\to2$ from the left drops dramatically \cite{ruschhaupt_2006_diode}.  
The role and importance of the state-selective mirror potentials is thus 
not reduced to blocking. Their presence is essential to facilitate an adiabatic 
rightwards transfer of population from state 1 to state 2 \cite{ruschhaupt_2006_diode}.  
This makes 
the device quite robust and insensitive to shape or parameter variations, in particular 
reflection and transmission probabilities become in 1D flat functions of the velocity
for broad intervals. In 3D, these intervals will in general also depend on the
incidence angle, but 
later we shall show that this dependence can be suppressed using an appropriate detuning.
     
For many applications, the atom should decay irreversibly back to the ground
state after crossing the three regions to avoid any backwards motion from
right to left. 
The decay can occur naturally or be forced 
by an additional quenching laser pumping the atom to a
third state which decays to the ground state with a fast rate $\gamma_3$,
see Fig. \ref{fig2}b. The quenching laser interaction is characterized
by a wave vector of modulus $k_Q>0$ in $y$ direction,
Rabi frequency  $\Omega_Q$,
and detuning $\Delta_Q$ (difference between the laser frequency
and the transition frequency $\omega_{32}$).

The atom dynamics corresponding to Fig. \ref{fig2} can be described by a
master equation or equivalently by the quantum jump formalism \cite{qj}. 
However, to determine the working conditions of the diode, it is enough to 
examine the evolution until the
first spontaneous decay of the atom,
since, with the chosen laser configuration, 
it is extremely unlikely that
there will be more than one spontaneous decay.
This evolution is described by an effective Schr\"odinger-equation dynamics
in the quantum jump approach.
Using the rotating-wave approximation, and in a laser adapted 
interaction picture such that the Hamiltonian
is time independent, the corresponding effective
Hamiltonian is 
\begin{eqnarray*}
\lefteqn{H_{2} =\frac{{p}_x^{\,2}}{2m}
+ \frac{{p}_y^{\,2}}{2m}} &&\\
& & +
\frac{\hbar}{2}\left(
\begin{array}{ccc}
W_1 (x) & \Omega_P (x) e^{-ik_P y} & 0\\
\Omega_P(x) e^{ik_P y} & W_2 (x) - 2 \Delta_P & \Omega_Q (x) e^{-i k_Q y}\\
0 & \Omega_Q (x) e^{i k_Q y} & -i \gamma_3 - 2 (\Delta_Q + \Delta_P)
\end{array}\right),
\end{eqnarray*}
with 
$|1\rangle \equiv \left(\begin{array}{c}1\\0\\0 \end{array}\right)$,
$|2\rangle \equiv \left(\begin{array}{c}0\\1\\0 \end{array}\right)$, and
$|3\rangle \equiv \left(\begin{array}{c}0\\0\\1 \end{array}\right)$.
${p}_x = -i\hbar \frac{\partial}{\partial x}$
and ${p}_y = -i\hbar \frac{\partial}{\partial y}$ are
the momentum operator components in $x-y$ space, and $m$ is the mass
(corresponding to Neon in all numerical examples).
$W_1(x)\hbar/2$ and $W_2(x)\hbar/2$ are the effective reflecting
potentials. Their dependence on $x$ only (and not on $y$) is justified
in the Appendix for a three-laser realization of the 2-level diode, and in 
section III for a two-laser realization.  
We have $k_P = k_{P0} + \frac{\Delta_P}{c}$
and $k_Q = k_{Q0} + \frac{\Delta_Q}{c}$ with
$k_{P0} = \frac{\omega_{21}}{c}$
and $k_{Q0} = \frac{\omega_{32}}{c}$
(we assume $\omega_3 > \omega_2 > \omega_1$).

We will examine the stationary Schr\"odinger equation
\begin{eqnarray}
E \Psi (x,y) = H_{2} \Psi (x,y)
\label{main2l}
\end{eqnarray}
with $E = \frac{\hbar^2}{2m} \left( k_x^2 + k_y^2 \right)$,
$k_x := \frac{m}{\hbar} v\, \cos\Theta \ge 0$,
$k_y := \frac{m}{\hbar} v\, \sin\Theta$, and
$v=\fabs{\vec{v}}$ (see Fig. \ref{fig2}d). 
To disentangle different effects we shall 
first analyze the diode part shown
in Fig. \ref{fig2}a and determine the optimal working conditions for which 
atoms coming from the right are reflected and atoms from the left are
transmitted.
Then we shall study the quenching laser alone, as in Fig.\ref{fig2}b, 
and finally combine everything.

\subsection{Effective one-dimensional equations}

The two-dimensional
equation (\ref{main2l}) can be transformed into an
effective one-dimensional one by inserting the ansatz
\begin{eqnarray*}
\Psi (x,y) = \left(\begin{array}{l}
\phi_1(x) e^{i k_y y}\\
\phi_2(x) e^{i (k_y + k_P) y}\\
\phi_3(x) e^{i (k_y + k_P + k_Q) y}  
\end{array}\right).
\end{eqnarray*}
Then we get
\begin{widetext}
\begin{eqnarray}
\frac{\hbar^2 k_x^2}{2m} 
\left(\begin{array}{c}\phi_1(x)\\ \phi_2(x)\\ \phi_3(x)\end{array}\right) =
\Bigg[\frac{{p}_x^2}{2m} +
 \frac{\hbar}{2} \left(\begin{array}{ccc}
W_1(x) & \Omega_P(x) & 0\\
\Omega_P (x) & W_2 (x) - 2\Delta_{3d,2} & \Omega_Q (x)\\
0 & \Omega_Q (x) & -2(\Delta_{3d,2}+\Delta_{3d,3})-i \gamma_3
\end{array}\right)\Bigg]
\left(\begin{array}{c}\phi_1(x)\\ \phi_2(x)\\ \phi_3(x)\end{array}\right),
\label{eff1d}
\end{eqnarray}
\end{widetext}
where
\begin{eqnarray*}
\Delta_{3d,2} &=& \Delta_P-\frac{\hbar}{2m}(2 k_y k_P + k_P^2)\\
\Delta_{3d,3} &=& \Delta_Q -\frac{\hbar}{2m}[2 (k_y+k_P) k_Q + k_Q^2]
\end{eqnarray*}
are effective ``three-dimensional'' detunings which include 
Doppler and recoil terms.

\subsection{Diode without quenching}

First we shall study the part of
the atom diode shown in Fig. \ref{fig2}a by 
setting $\Omega_Q = 0$, $\gamma_3 = 0$.
Let us consider left incidence in the ground
state. Then the asymptotic form of $\Psi$ is
\begin{eqnarray*}
\Psi (x,y) &\stackrel{x\ll0}{=}&
\left(\begin{array}{l}
e^{i k_x x + i k_y y}\\
0\\
0
\end{array}\right)\\ & & +
\left(\begin{array}{l}
R_{11} e^{-i k_x x + i k_y y}\\
R_{21} e^{-i q x + i (k_y + k_P) y}\\
0
\end{array}\right),\\
\Psi (x,y) &\stackrel{x\gg0}{=}&
\left(\begin{array}{l}
T_{11} e^{i k_x x + i k_y y}\\
T_{21} e^{i q x + i (k_y + k_P) y}\\
0
\end{array}\right),
\end{eqnarray*}
with
\begin{eqnarray}
q &=& \sqrt{k_x^2 - k_P^2 - 2k_y k_P + 2m\Delta_P/\hbar}
\\
\nonumber
&=& \frac{m}{\hbar} \sqrt{v^2 \cos^2\Theta - v_P^2 - 2 v_P v \sin\Theta
+ 2c\Delta v_P},
\label{defq}
\end{eqnarray}
where $v_P = v_{P0} + \Delta v_P$,
$v_{P0} = \hbar k_{P0}/m$, and $\Delta v_P = \hbar\Delta_P/(mc)$.
We denote by $R_{\beta\alpha}=R_{\beta\alpha} (w>0,\Theta)$
the reflection amplitudes for incidence with modulus of velocity $w=v>0$
and incidence angle $\Theta$ from the left in channel $\alpha$,
and reflection in channel $\beta$.  
Similarly we denote 
by $T_{\beta\alpha}=T_{\beta\alpha} (w>0,\Theta)$ the transmission 
amplitudes for incidence in channel $\alpha$ from
the left and transmission in channel $\beta$.
The reflection and transmission coefficients can be simply
calculated for the effective one-dimensional
equation (\ref{eff1d}) with the effective ``three-dimensional'' detunings.  
Perfect ``diodic'' behavior means
full transmission for left incidence, i.e.
\begin{eqnarray}
\frac{\mbox{Re} (q)}{k_x} \fabsq{T_{21}} \approx 1,
\label{eqtrans}
\end{eqnarray}
such that the approximate, asymptotic form of $\Psi$ is
\begin{eqnarray}
\Psi (x,y) &\stackrel{x\ll0}{\approx}&
\left(\begin{array}{l}
e^{i k_x x + i k_y y}\\
0\\
0
\end{array}\right),\\
\Psi (x,y) &\stackrel{x\gg0}{\approx}&
\left(\begin{array}{l}
0\\
T_{21} e^{i q x + i (k_y + k_P) y}\\
0
\end{array}\right).\label{psiout}
\end{eqnarray}
The important point is that, for having a traveling (and no evanescent)
transmitted wave and for fulfilling Eq. (\ref{eqtrans}), 
$q$ must be real.
From this condition we can derive a lower boundary for $k_x$
at an incidence angle $\Theta$, namely
\begin{eqnarray}
& & k_x^2 \ge k_{L}^2 + 2 k_y k_{L} - 2m\Delta_P/\hbar.
\label{xxx}
\end{eqnarray}
Note that atomic incidence opposite to the laser direction is more 
efficient than incidence in the same direction because of the Doppler effect. 
In this later case ($k_y>0$), a positive (blue) detuning may compensate the
effect of Doppler
and recoil terms.  
Because the detuning is much smaller than the transition frequency  
$\omega_{21}$, $k_P \approx k_{P0}$. So
we can write the condition (\ref{xxx}) as 
\begin{eqnarray}
v^2 \cos^2\Theta
- 2 v_{P0} v \sin\Theta  + 2c \Delta v_P - v_{P0}^2 \ge 0.
\label{lbound2l}
\end{eqnarray}

Let us now consider incidence from the right in the ground state.
The asymptotic form for the wave function is then
\begin{eqnarray*}
\Psi (x,y) &\stackrel{x\gg0}{=}&
\left(\begin{array}{l}
e^{-i k_x x + i k_y y}\\
0\\
0
\end{array}\right)\\
& & +
\left(\begin{array}{l}
R_{11} e^{i k_x x + i k_y y}\\
R_{21} e^{i q x + i (k_y + k_P) y}\\
0
\end{array}\right),\\
\Psi (x,y) &\stackrel{x\ll0}{=}&
\left(\begin{array}{l}
T_{11} e^{-i k_x x + i k_y y}\\
T_{21} e^{-i q x + i (k_y + k_P) y}\\
0
\end{array}\right),\\
\end{eqnarray*}
where the right
incidence will be indicated, if necessary, with a negative argument $w$ in
transmission and reflection amplitudes, $w=-v<0$.  
Perfect ``diodic'' behavior means now full reflection, i.e.
$\fabsq{R_{11}} \approx 1$. An  approximate condition for this to
happen is that $\frac{\hbar^2}{2m} k_x^2 \le \frac{\hbar}{2} \widehat{W}_1$,
where $\widehat{W}_1 = \mbox{max}_x W_1(x)$, and therefore we get
a velocity bound for full reflection, namely
\begin{eqnarray}
v \quad \le \quad 
\sqrt{\frac{\hbar\widehat{W}_1}{m}} \frac{1}{\cos\Theta}\, .
\label{rbound}
\end{eqnarray}
Note that this bound
is symmetrical with respect to normal incidence because the 
mirror barrier height, by hypothesis, does not depend on the incident angle, 
see the Appendix for the physical justification. 
 
%
\begin{figure}[t]
\begin{center}
\includegraphics[angle=0,width=\linewidth]{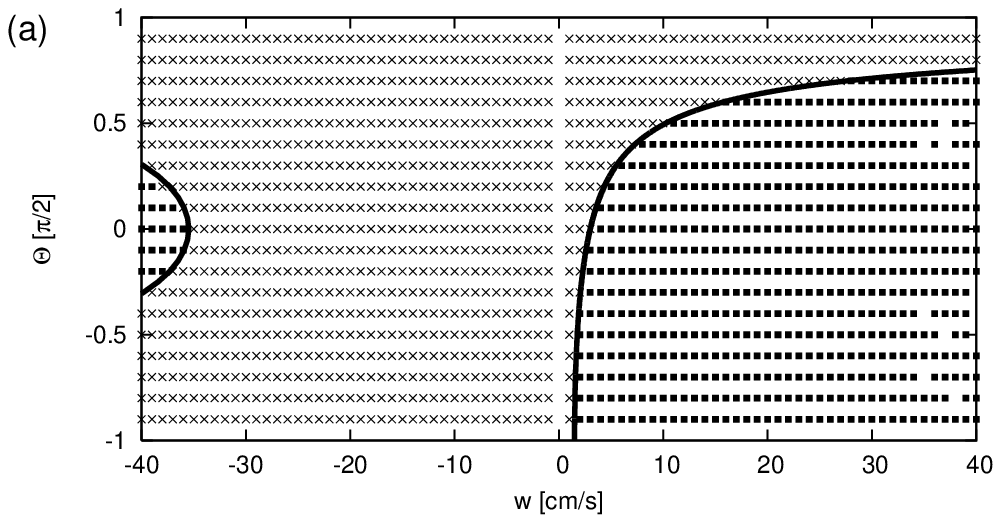}

\includegraphics[angle=0,width=\linewidth]{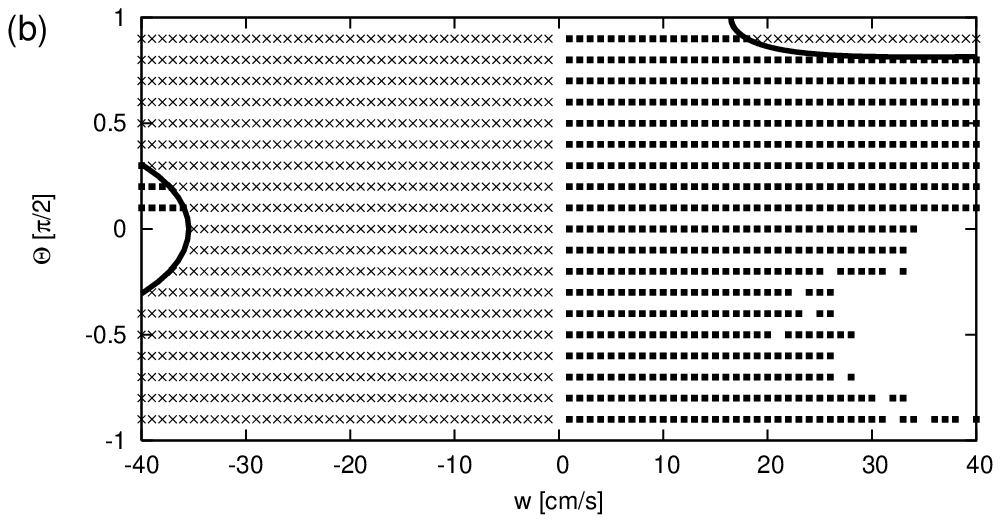}
\end{center}
\caption{\label{fig3}Diodic behavior based on 2 levels without quenching: 
full reflexion (crosses),
full transmission (boxes),
otherwise (blanks);
$d=50\mum$, $\sigma=15\mum$, $\widehat{\Omega}_P = 10^6\si$,
$\widehat{W}_1 = \widehat{W}_2 = 4\times 10^7\si$, $v_{P0}=3 \cms$;
(a) $\Delta v_P = 0$;
(b) $\Delta v_P = 1.8\times 10^{-9}\cms$;
the solid line on the left indicates the boundary following from Eq.
(\ref{rbound}) and the one on the right the boundary
resulting from Eq. (\ref{lbound2l}).}
\end{figure}
%

A numerical example can be found in Fig. \ref{fig3}.
The shapes of the Rabi frequency and the reflecting potentials 
are Gaussian, 
$\Omega_P (x)= \widehat{\Omega}_P \;\Pi(x, 0)$,
$W_1 (x) = \widehat{W}_1 \;\Pi(x, d)$, and
$W_2 (x) = \widehat{W}_2 \;\Pi(x, -d)$,
with
\begin{eqnarray}
\Pi (x, x_0)&=& \fexp{-\frac{(x-x_0)^2}{2 \sigma^2}}.
\label{gaussian}
\end{eqnarray}
Full reflexion is indicated with crosses, more precisely
this means 
\begin{eqnarray}   
\lefteqn{\left(1.0 - |R_{11}|^2\right) + \frac{\mbox{Re} (q)}{k_x}
  |R_{21}|^2} & &
\nonumber\\
& & + 
|T_{11}|^2 + \frac{\mbox{Re} (q)}{k_x} |T_{21}|^2  < 0.01.
\label{fullref}
\end{eqnarray}
Full transmission is indicated by boxes, this means
\begin{eqnarray}
\lefteqn{|R_{11}|^2 + \frac{\mbox{Re} (q)}{k_x} |R_{21}|^2} &&
\nonumber\\
&& + |T_{11}|^2 +
\left(1.0 - \frac{\mbox{Re} (q)}{k_x} |T_{21}|^2\right) < 0.01, 
\label{fulltrans}
\end{eqnarray}
an any other result is left blank.

Let us first discuss the case of incidence from the right ($w < 0$): 
We can see in Figs. \ref{fig3} that the breakdown of the full
reflexion is very well approximated by the boundary (\ref{rbound})
(left solid line). The diode is fully reflecting the 
ground state atoms independently  of the
incident angle and detuning for $v\lesssim 35 \cms$.

For left incidence, $w > 0$, 
in Fig. \ref{fig3}a with an on-resonance
pumping laser,  the range of the
diodic behavior is not independent of the angle, and there is 
no $v_{max}$ such that the atoms incident from the left are
transmitted for all velocities $w\le v_{max}$
independently of the angle of incidence.
The minimal velocity for transmission can be approximated
by the boundary (\ref{lbound2l}) (see the right solid line
in the figure). As shown in Fig. \ref{fig3}b,
full transmission for the lowest velocities 
can be realized independently of the angle by using detuning $\Delta_P$:
now the atom diode is working for
$v\le v_{max} \approx 16 \cms$ for left and right
incidence independently of the angle of incidence.
Increasing the detuning, $v_{max}$ cannot be
increased to arbitrarily high values, because of the
earlier breakdown of the
adiabatic transmission (see the blanks in Fig. \ref{fig3}b).
Nevertheless, the breakdown velocity of the
adiabatic transmission can be increased by increasing
the laser intensities.

\subsection{Quenching}

\begin{figure}[t]
\begin{center}
\includegraphics[angle=0,width=0.8\linewidth]{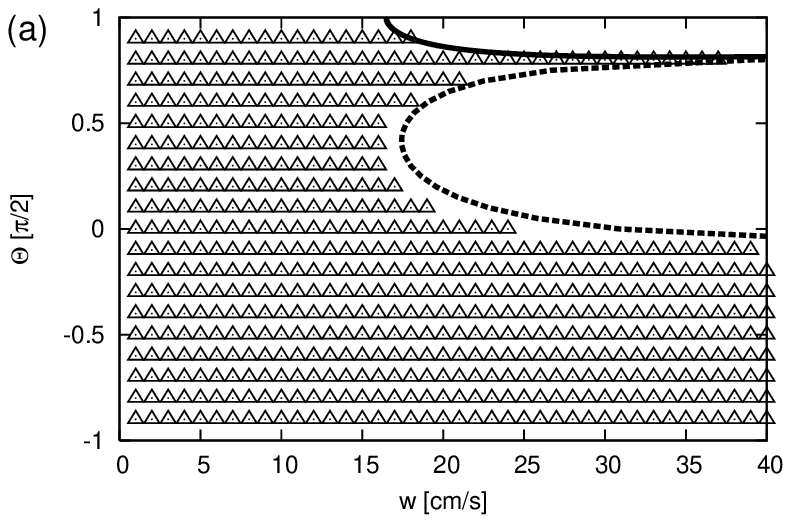}

\includegraphics[angle=0,width=0.8\linewidth]{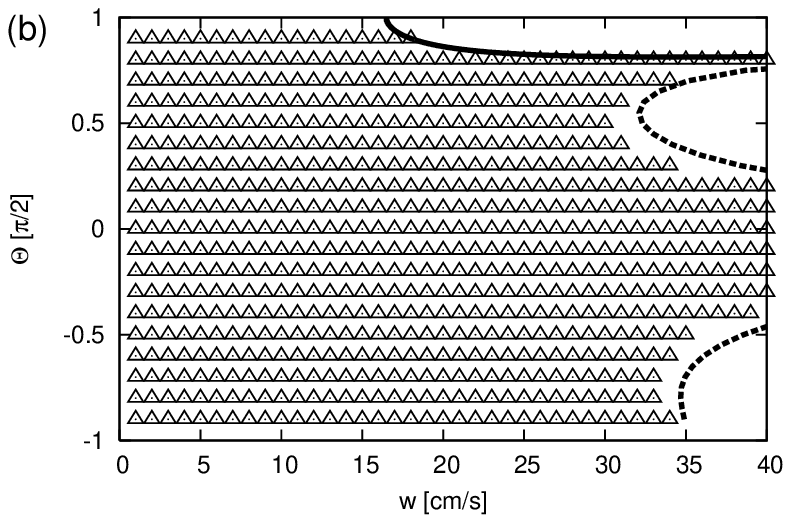}

\includegraphics[angle=0,width=0.8\linewidth]{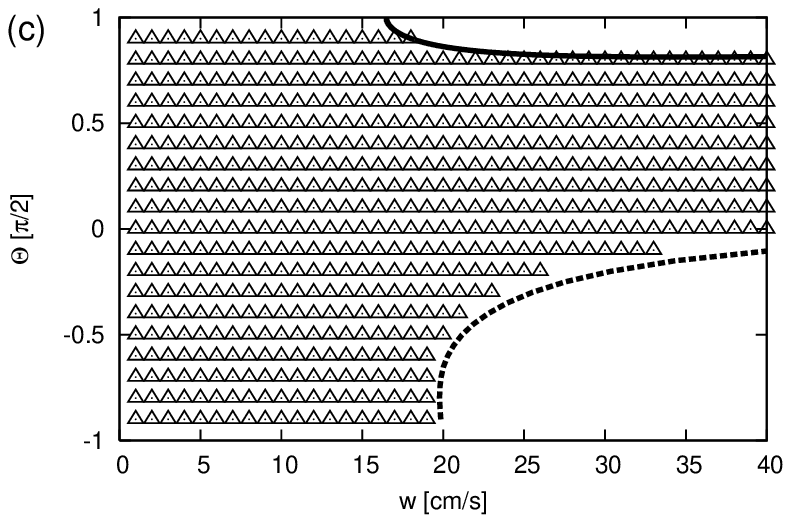}
\end{center}
\caption{\label{fig4}
Quenching: the triangles indicate full absorption;
$\widehat{\Omega}_Q = 2 \times 10^6\si$, 
$v_{Q0} = 3 \cms$, $\gamma_3 = 3\times 10^5\si$,
(a) $\Delta v_Q = -1.8\times 10^{-9} \cms$;
(b) $\Delta v_Q = 0$;
(c) $\Delta v_Q = 1.8\times 10^{-9} \cms$;
for other parameters see Fig. \ref{fig3}b.}
\end{figure}

We shall now consider the quenching laser alone, as in Fig. \ref{fig2}b,
i.e. we set $\Omega_P = 0$, $W_1 = 0$, and $W_2 = 0$, but
$\Omega_Q > 0$ and $\gamma_3 > 0$. Also, we
restrict ourselves to the case of incidence from the left.
Motivated by the full transmission case of the last subsection,
see Eq. (\ref{psiout}),
we assume that the asymptotic forms are
\begin{eqnarray*}
\Psi (x,y) &\stackrel{x\ll0}{=}&
\left(\begin{array}{l}
0\\
e^{i q x + i (k_y + k_P) y}\\
0
\end{array}\right)\\ & & +
\left(\begin{array}{l}
0\\
R_{22} e^{-i q x + i (k_y + k_P) y}\\
R_{32} e^{-i q' x + i (k_y + k_P + k_Q) y}
\end{array}\right),\\
\Psi (x,y) &\stackrel{x\gg0}{=}&
\left(\begin{array}{l}
0\\
T_{22} e^{i q x + i (k_y + k_P) y}\\
T_{32} e^{i q' x + i (k_y + k_P + k_Q) y}
\end{array}\right),
\end{eqnarray*}
with $q = q(w,\Theta)$ given by Eq. (\ref{defq}) and
$q'   = \sqrt{q^2 - k_Q^2 - 2 (k_y+k_P) k_Q + 2m\Delta_Q/\hbar}$.
Note that we have included an incident wavenumber $k_P$ in $y$ direction
since we assume that the atom has previously passed successfully the diode 
of Fig. \ref{fig2}a.
A spontaneous decay should happen with high probability, therefore
the goal is now nearly full absorption, i.e.
all reflection and transmission coefficients should vanish.
We have examined this numerically in Fig. \ref{fig4}
where $v_{Q0} = \hbar k_{Q0}/m$ and $\Delta v_Q = \hbar\Delta_Q/(mc)$.
The triangles mark velocities $w=v>0$ and angles $\Theta$
where we get practically full absorption, i.e.
$ \fabsq{R_{22}}
+ \frac{\mbox{Re} (q')}{q} \fabsq{R_{32}}
+ \fabsq{T_{22}}
+ \frac{\mbox{Re} (q')}{q} \fabsq{T_{32}} < 0.01$. We get practically
full absorption independent of the incidence angle $\Theta$ for
$w \lesssim 16 \cms$. 

Two different reasons for failure may be identified: in the upper part of the
figure,  the solid line is the bound for evanescent waves in
Eq. (\ref{lbound2l}). In the
surrounded blank region there is no wave with real momentum incident on the
quenching potential, i.e. $q$ is imaginary. The other blank regions are due to
the failure of absorption with $T_{22}\neq 0$ because of a weak quenching. 
A simple understanding of the boundary shapes found follows
from a complex potential approximation.
By adiabatic elimination, as in the Appendix, a Schr\"odinger
equation for level 2 of (\ref{eff1d}) can be written in terms of a complex
potential $W_{eff}\hbar/2$, 
\begin{eqnarray*}
\lefteqn{\frac{\hbar^2 k_x^2}{2m} \phi_2(x)
= \frac{{p}_x^2}{2m}\phi_2 (x) +} &&\\
&& \frac{\hbar}{2}\left[\Omega_P (x) \phi_1(x) + \left(W_2 (x) - 2\Delta_{3d,2}
 +W_{eff} (x)\right) \phi_2 (x)\right]
\end{eqnarray*}
with
\begin{eqnarray*}
W_{eff} (x)
&=& \frac{2 \Delta_{3d,3}\,\Omega_Q(x)^2}{4\Delta_{3d,3}^2 + \gamma_3^2} - i
\underbrace{\frac{\gamma_3\, \Omega_Q(x)^2}{4\Delta_{3d,3}^2 +
    \gamma_3^2}}_{\gamma_{eff} (x)}\, .
\end{eqnarray*}

Absorption will fail when the crossing time $\sim \frac{\sigma}{v_x}$
($\sigma$ being the width of the quenching laser, 
see Eq.(\ref{gaussian}))
is lower than the corresponding lifetime $\sim 1/\gamma_{eff}$, i.e.
an approximate condition for absorption is
\begin{eqnarray*}
\frac{\gamma\,\widehat{\Omega}_Q^2}{4\Delta_{3d,3}^2 + \gamma_3^2}
\cdot \frac{\sigma}{v_x} > \beta
\end{eqnarray*}
with $\beta$ being a fitting constant. The boundaries following from
this condition with $\beta=3$ are plotted in Fig. \ref{fig4} by dashed lines.
For negative (see. Fig. \ref{fig4}a)/positive (see. Fig. \ref{fig4}c)
detuning, $\gamma_{eff}$ is maximal at
negative/positive angles, respectively, and decreases otherwise. The effect of
small absorption, and large lifetime, is however compensated for very oblique
incidence, so that the crossing time is even larger than the lifetime: this
explains the roughly parabolic profiles of the dashed lines.
The absorption failure regions will shrink by increasing laser intensity.  
The zero detuning case is globally advantageous and we shall restrict to it
in the following.

%
%

\subsection{Diode with quenching}

\begin{figure}
\begin{center}
\includegraphics[angle=0,width=\linewidth]{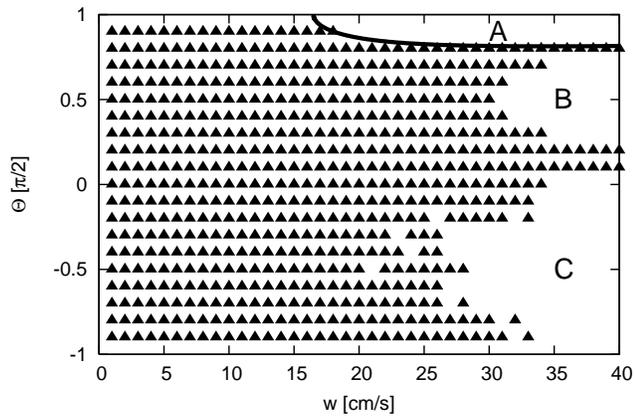}
\end{center}
\caption{\label{fig5}Diodic behavior based on 2 levels plus
quenching: the filled triangles
indicate full transmission to the second state without quenching laser
and full absorption if a quenching potential is added;
the characters indicate regions of different breakdown reasons as explained in the main text;
parameters in Figs. \ref{fig3}b and \ref{fig4}b.}
\end{figure}

Now we are ready to combine the diode and the quenching laser as in
Fig. \ref{fig2}c and we restrict again the analysis to incidence from the left.
We use the parameters of Fig. \ref{fig3}b and Fig. \ref{fig4}b and
the result is plotted in Fig. \ref{fig5}:
the filled triangles mark velocities and angles for which we get
full transmission through the first three regions without quenching laser and full
absorption if the quenching laser is switched on, i.e.
the atom diode is working: because with the chosen laser configuration, 
it is extremely likely that the excited atom will decay in the quenching
region after it has passed
the mirror potential $W_2 \hbar/2$ and therefore it will move finally
to the right-hand side. 
This happens for velocities $w \lesssim 16
\cms$ independently  of the angle of incidence $\Theta$.    
The breakdown at high velocities
in Fig. \ref{fig5} has different reasons: in the
blank region~A (bordered by the solid line)
there is full reflection instead of transmission
to state $2$; in the
blank region~B the quenching process fails
(compare to Fig. \ref{fig4}b); and in the blank region~C 
the pumping process to state $2$ fails, compare to Fig. \ref{fig3}b. 

%
%

%
%
%
%
%
%
%
%
%
\section{Atom diode based on $3$ levels plus quenching}
%
%
%
%
%
%
\begin{figure}
\begin{center}
\vspace*{0.5cm}

\includegraphics[angle=0,width=0.9\linewidth]{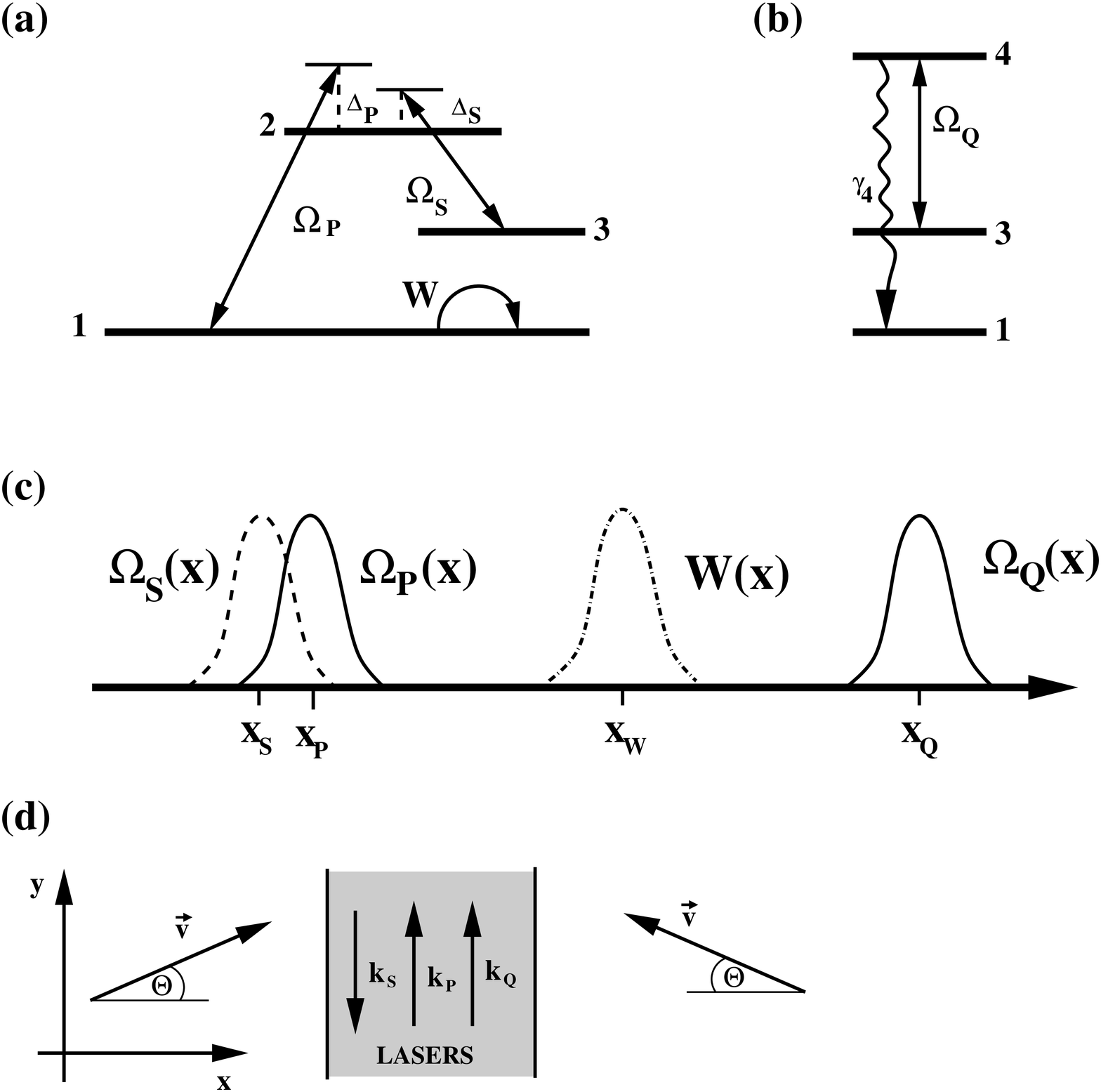}
\end{center}
\caption{\label{fig6}Atom diode based on $3$ levels plus quenching:
(a,b) Schematic action of the different lasers 
on the atom levels, 
(c) location of the different laser potentials in $x$ direction, and
(d) setting in the $x-y$ plane.}
\end{figure}
%

An atom diode may also be realized with the three level scheme 
shown in Figs. \ref{fig5}a and \ref{fig5}c.
The basic idea is to combine two lasers that achieve STIRAP (stimulated  
Raman adiabatic passage) with a state-selective 
reflecting interaction for the ground state.
The STIRAP method is well known 
\cite{bergmann.1998} and consists of an adiabatic transfer of population 
between levels 1 and 3 by two partially overlapping
(here in space, see Fig. \ref{fig5}c)
laser beams coupling 1 and 3 with a further level 2,
in a somewhat counterintuitive order.  
The pump laser couples the atomic levels 1 and 2 with
Rabi frequency $\Omega_P$, and the Stokes laser couples the states 2 and 3
with Rabi frequency $\Omega_S$.
Because by a STIRAP transfer the atom is (ideally) transfered
from $1\to 3$ without being in $2$ the transfer is
essentially not affected by a possible decay $2\to 1$.
If we add a third laser causing an 
effective reflecting potential $W(x)\hbar/2$ 
for the ground state component,  
an atom incident from the right in the ground state
and with low enough velocity is  
reflected by the potential $W (x)\hbar/2$
and will finally travel to the right.

If the atom comes 
from the left in the ground state, it will be 
transfered by STIRAP to the third state.
Therefore the moving atom in the third state
is not affected by $W (x)$. If quenching is included, see Fig. \ref{fig5}b,
it will be pumped to a fourth level by the quenching laser and 
decay to the ground state. Another possibility is to pump
it to the second level again if this level decays fast to the
ground state \cite{ruschhaupt_2006_quench}. Finally the atom
will travel to the right in the ground state.

Analogously to the previous section,
it is sufficient to examine an effective stationary Schr\"odinger equation
which takes now the form
\begin{eqnarray}
E \Psi (x,y) =  H_{3} \Psi (x,y)
\label{main3l}
\end{eqnarray}
with $E = \frac{\hbar^2}{2m} \left( k_x^2 + k_y^2 \right)$,
$k_x := \frac{m}{\hbar} v\, \cos\Theta \ge 0$,
$k_y := \frac{m}{\hbar} v\, \sin\Theta$, and
$v=\fabs{\vec{v}}$ (see Fig. \ref{fig6}d).
The effective Hamiltonian is given by 
\begin{widetext}
\begin{eqnarray*}
H_{3} = \frac{{p}_x^{\,2}}{2m} 
+ \frac{{p}_y^{\,2}}{2m} +
\frac{\hbar}{2} \left(\begin{array}{cccc}
W (x) & \Omega_P (x) e^{-i k_P y} & 0 & 0\\
\Omega_P (x) e^{i k_P y} & -2\Delta_P & \Omega_S (x) e^{i k_S y} & 0\\
0 & \Omega_S (x) e^{-i k_S y} & -2\Delta_P+2\Delta_S 
& \Omega_Q (x) e^{-i k_Q y}\\
0 & 0 & \Omega_Q (x) e^{i k_Q y} & -2\Delta_P+2\Delta_S-i\gamma_4
\end{array}\right).
\end{eqnarray*}
\end{widetext}
$k_P$ is chosen similarly to the previous section.
The Stokes laser has a wave vector of modulus $k_S > 0$
in $-y$ direction. We get
$k_S = k_{S0} + \frac{\Delta_S}{c}$ with
$k_{S0} = \frac{\omega_2 - \omega_3}{c}$.
The on-resonance quenching laser has a wave vector
in $y$ direction
of modulus $k_Q = \frac{\omega_4 - \omega_3}{c}$.
Note that we assume for the frequencies of the atomic levels
$\omega_4 > \omega_2 > \omega_3 > \omega_1$.

\subsection{Reduction to the 2 level plus quenching case}

This level and laser scheme
without mirror potential, i.e. $W (x) = 0$, can be used
to realize the two-level case and potentials of the previous section
in the limit of highly detuned lasers.
In \cite{ruschhaupt_2006_diode} we have discussed a similar result
in the one-dimensional case. 
We shall show now that this reduction is also valid in
three dimensions and including a final detuning in the effective 
pumping interaction.  
Let $\Psi = (\psi_1,\psi_2,\psi_3,\psi_4)^T$.
Equation (\ref{main3l}) for the second component $\psi_2$ with
$W (x) = 0$ is given by
\begin{eqnarray*}
\lefteqn{\left[E - \left(\frac{{p}_x^2}{2m} + \frac{{p}_y^2}{2m}\right)
+ \hbar\Delta_P\right] \psi_2 (x,y) =} & &\\
&&  \frac{\hbar}{2} \Omega_P (x) e^{i k_P y} \psi_1 (x,y)
+ \frac{\hbar}{2} \Omega_S (x) e^{i k_S y} \psi_3 (x,y)\, .
\end{eqnarray*}
In the case of large detuning $\Delta_P$
we may write heuristically
\begin{eqnarray*}
\psi_2 (x,y) &\approx&
\frac{1}{2\Delta_P}
\left[\Omega_P (x) e^{i k_P y} \psi_1 (x,y)\right.\\
& & \left. + \Omega_S (x) e^{i k_S y} \psi_3 (x,y)\right]. 
\end{eqnarray*}
We assume that $\Delta_S$ is also large such that
$\Delta_P - \Delta_S$ is ``small''.
Putting the approximate expression for $\psi_2$ in the equations
for the other components we get
\begin{widetext}
\begin{eqnarray*}
E\,
\left(\begin{array}{c}
\psi_1 (x,y)\\
\psi_3 (x,y)\\
\psi_4 (x,y)
\end{array}\right) =
\left[\frac{{p}_x^2}{2m} + \frac{{p}_y^2}{2m} +
\frac{\hbar}{2} \left(\begin{array}{ccc}
\widetilde{W}_1 (x) & \widetilde{\Omega} (x) e^{-i \tilde{k}_L y} & 0\\
\widetilde{\Omega} (x) e^{i \tilde{k}_L y} &
\widetilde{W}_3 (x)-2\widetilde{\Delta} &
\Omega_Q (x) e^{-i k_Q y}\\
0 & \Omega_Q (x) e^{i k_Q y} & -2\widetilde{\Delta}-i\gamma_4
\end{array}\right)\right]
\left(\begin{array}{c}
\psi_1 (x,y)\\
\psi_3 (x,y)\\
\psi_4 (x,y)
\end{array}\right)
\end{eqnarray*}
with
\begin{eqnarray*}
\widetilde{W}_1(x) = \frac{\Omega_P (x)^2}{2\Delta_P} , \;
\widetilde{W}_3(x) = \frac{\Omega_S (x)^2}{2\Delta_P}, \;
\widetilde{\Omega} (x) = \frac{\Omega_P (x) \Omega_S (x)}{2\Delta_P},\;
\tilde{k}_L = k_P - k_S ,\;
\widetilde{\Delta}  =  \Delta_P-\Delta_S,
\end{eqnarray*}
which has the same structure as Eq. (\ref{main2l}). The two lasers 
provide state-selective reflecting mirrors in the extremes, and a  
pumping interaction in their overlapping region with an effective 
wavenumber and detuning equal to the differences 
between wavenumbers and detunings.     

\subsection{Effective one-dimensional equation}

Now we return to Eq. (\ref{main3l}). Again we can rewrite the equation
as an effective one-dimensional equation by inserting the ansatz
\begin{eqnarray*}
\Psi (x,y) = \left(\begin{array}{l}
\phi_1(x) e^{i k_y y} \\
\phi_2(x) e^{i (k_y + k_P) y} \\
\phi_3(x) e^{i (k_y + k_P - k_S) y}\\ 
\phi_4(x) e^{i (k_y + k_P - k_S + k_Q) y}\\ 
\end{array}\right)
\end{eqnarray*}
to get
\begin{eqnarray*}
\frac{\hbar^2 k_x^2}{2m} 
\left(\begin{array}{c}\phi_1(x)\\ \phi_2(x)\\ \phi_3(x)\\\phi_4(x)
\end{array}\right) =
 \left[\frac{p_x^2}{2m} + 
\frac{\hbar}{2} \left(\begin{array}{cccc}
W (x) & \Omega_P (x) & 0 & 0\\
\Omega_P (x) & -2 \Delta_{3d,2} & \Omega_S (x) & 0\\
0 & \Omega_S (x) & -2 (\Delta_{3d,2}+\Delta_{3d,3}) & \Omega_Q (x)\\
0 & 0 & \Omega_Q (x) & -2 (\Delta_{3d,2}+\Delta_{3d,3}+\Delta_{3d,4})
- i \gamma_4
\end{array}\right)\right]
\left(\begin{array}{c}\phi_1(x)\\ \phi_2(x)\\ \phi_3(x)\\ \phi_4(x)
\end{array}\right),
\end{eqnarray*}
\end{widetext}
where the effective ``three-dimensional'' detunings
are defined by 
\begin{eqnarray*}
\Delta_{3d,2} &=& \Delta_P-\frac{\hbar}{2m}(2 k_y k_P + k_P^2)\, ,\\
\Delta_{3d,3} &=& - \Delta_S
-\frac{\hbar}{2m}(-2 (k_y+k_P) k_S + k_S^2)\, ,\\
\Delta_{3d,4} &=& -\frac{\hbar}{2m}(2 (k_y+k_P-k_S) k_Q + k_Q^2)\, .
\end{eqnarray*}

\subsection{Diodic behavior}

Similarly to the previous section we first want to derive some
boundary conditions for the diodic behavior.
For right incidence, we get the same upper boundary (\ref{rbound})
for full reflection. 
For left incidence in the ground
state and without a quenching laser ($\Omega_Q=0,\gamma_4=0$)
the asymptotic form is
\begin{eqnarray*}
\Psi (x,y) &\stackrel{x\ll0}{=}&
\left(\begin{array}{l}
e^{i k_x x + i k_y y }\\
0\\
0\\
0
\end{array}\right)\\ & & +
\left(\begin{array}{l}
R_{11} e^{-i k_x x + i k_y y }\\
R_{21} e^{-i q x + i (k_y + k_P) y }\\
R_{31} e^{-i q'' x + i (k_y + k_P - k_S) y }\\
0
\end{array}\right),\\
\Psi (x,y) &\stackrel{x\gg0}{=}&
\left(\begin{array}{l}
T_{11} e^{i k_x x + i k_y y }\\
T_{21} e^{i q x + i (k_y + k_P) y }\\
T_{31} e^{i q'' x + i (k_y + k_P - k_S) y }\\
0
\end{array}\right),
\end{eqnarray*}
with 
$k_x = \frac{m}{\hbar} v \cos\Theta \ge 0$,
$k_y = \frac{m}{\hbar} v \sin\Theta$,
$q = \sqrt{k_x^2 - k_P^2 - 2k_y k_P + 2m\Delta_P/\hbar}$, and
$q'' = \sqrt{k_x^2 - (k_P-k_S)^2 - 2k_y (k_P-k_S) 
+ 2m(\Delta_P-\Delta_S)/\hbar}$.
Following the same arguments as in the previous section,
we get a lower boundary for perfect transmission in
state $3$ (i.e. $\frac{\mbox{Re} (q'')}{k_x} \fabsq{T_{31}^l} \approx 1$)
for left incidence, namely
\begin{eqnarray}
v^2 \cos^2\Theta
- 2 v \sin\Theta v_{PS} + \Delta v_{PS} - v_{PS}^2 \ge 0,
\label{lbound3l}
\end{eqnarray}
where $v_{PS} = \hbar (k_{P0}-k_{S0})/m$ and
$\Delta v_{PS} = \hbar (\Delta_P-\Delta_S)/(mc)$.

%
\begin{figure}[t]
\begin{center}
\includegraphics[angle=0,width=\linewidth]{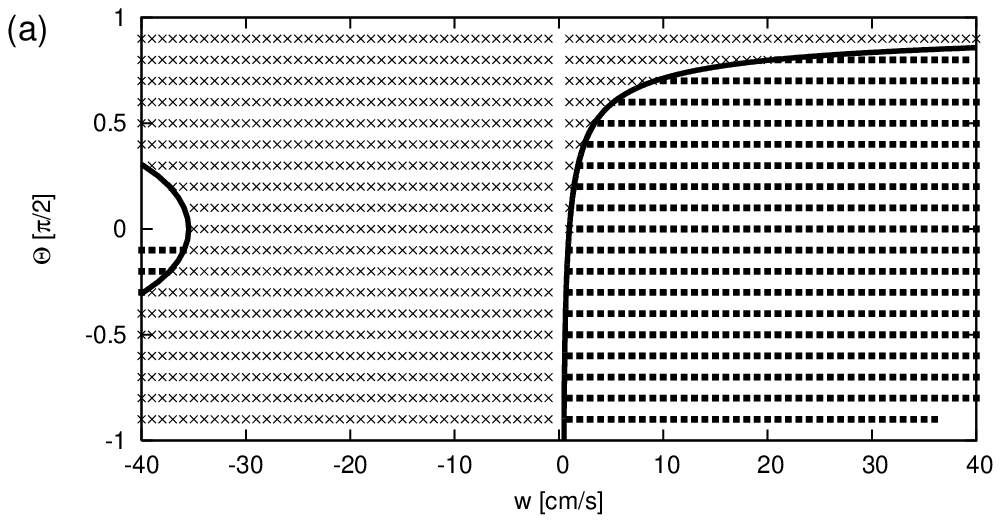}

\includegraphics[angle=0,width=\linewidth]{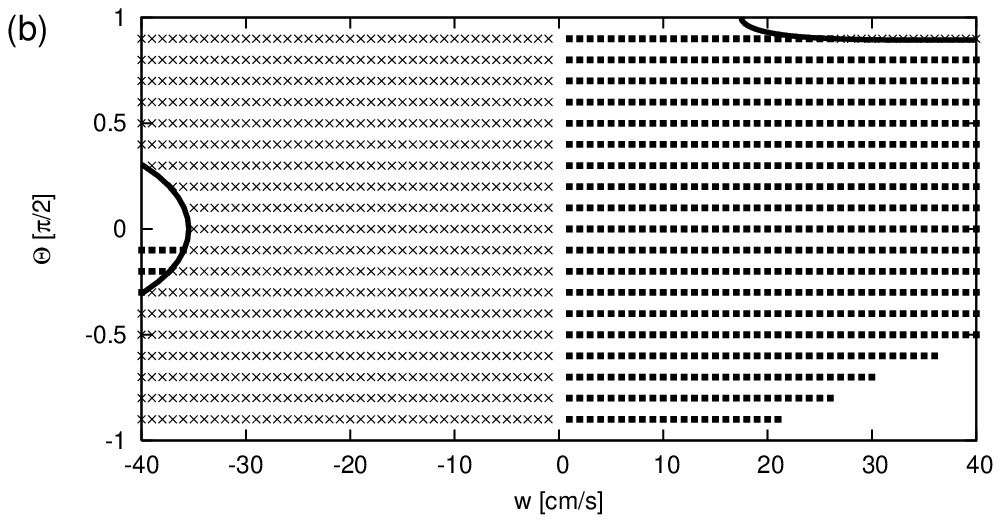}
\end{center}
\caption{\label{fig7}Diodic behavior based on 3 levels
without quenching: 
full reflexion (crossed), full transmission (boxes),
otherwise (blanks);
$\widehat{\Omega}_S = \widehat{\Omega}_P = 6\times 10^6\si$,
$\widehat{W} = 4\times 10^7\si$, $x_S=-20\mum$, $x_P=20\mum$,
$x_W=85\mum$, $v_{P0} = 3 \cms$, $v_{S0} = 2 \cms$, $\Delta_S=0$;
(a) $\Delta v_{PS} = 0$;
(b) $\Delta v_{PS} = 0.6 \times 10^{-9} \cms$;
the solid line on the left indicates the boundary following from Eq.
(\ref{rbound}) and the one on the right the boundary
resulting from Eq. (\ref{lbound3l}).}
\end{figure}
%

\begin{figure}[t]
\begin{center}
\includegraphics[angle=0,width=\linewidth]{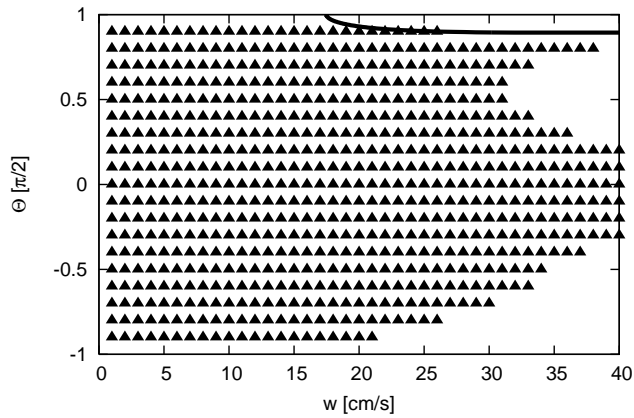}
\end{center}
\caption{\label{fig8}Diodic behavior based on 3 levels
plus quenching: the filled triangles
indicate full transmission to the third state without quenching laser
and full absorption if a quenching potential is added;
$v_{Q0} = 3 \cms$, $\Delta v_Q = 0$, 
$\widehat{\Omega}_Q = 2\times 10^6 \si$,
$x_Q= 150\mum$, $\gamma_4 = 3\times 10^5\si$, for 
other parameters see Fig. \ref{fig7}b.}
\end{figure}

A numerical example without quenching
can be found in Fig. \ref{fig7}.
The shapes of the Rabi frequency and the reflecting potentials 
in the model are again Gaussian, 
$\Omega_S (x)= \widehat{\Omega}_S \;\Pi(x, x_S)$,
$\Omega_P (x)= \widehat{\Omega}_P \;\Pi(x, x_P)$, and
$W (x) = \widehat{W} \;\Pi(x, x_W)$.
Full reflection (crosses) and full transmission to the third state (boxes)
are defined by a straightforward extension of the
condition (\ref{fullref}) resp. (\ref{fulltrans}).
Again, detuning is needed to fulfill the requirement
of a diode behavior independent of the angle, see Fig. \ref{fig7}b
for $v \lesssim 17 \cms$. 
Detuning causes STIRAP to break down earlier, as in Fig. \ref{fig3}b
for the two level model, nevertheless this breakdown can be 
pushed to higher velocities by increasing laser intensities.
The result if the quenching laser
($\Omega_Q (x)= \widehat{\Omega}_Q \;\Pi(x, x_Q)$)
is included
is shown in Fig. \ref{fig8} for incidence from the left.
The results are very similar to those of the two-level model,
the diode including quenching works for velocities 
$v \lesssim 17 \cms$ independently of the angle of incidence $\Theta$.

%
%

\section{Summary}
The prospects to realize in the laboratory atom diodes, i.e.
one-way laser barriers,
for selective trapping, cooling, implementing a Maxwell pressure demon, or
other applications to control the atomic motion, have motivated 
a series of theoretical studies to determine efficient atomic laser
configurations and internal level structure, as well as to identify
and quantify possible limitations.   

In this paper we have investigated three-dimensional effects in
an atom diode which
had been ignored in the existing
one-dimensional models. We have shown that the three-dimensional
case can be reduced to one-dimensional Schr\"odinger equations with 
``effective'' detunings taking into account Doppler and recoil effects. 
Simple formulae have been deduced for angle-dependent
upper and lower velocity bounds for correct ``diodic'' behavior.
The goal of a ``diodic'' behavior independent
of the incident angle up to a maximum velocity, a requirement to 
implement, for example, the Maxwell pressure demon, can be realized by properly
detuned lasers.

\begin{acknowledgments}
We acknowledge ``Acciones Integradas'' of the German
Academic Exchange Service (DAAD) and of Ministerio de
Educaci\'on y Ciencia (MEC). 
This work has also been supported by MEC through the project
FIS2006-10268-C03-01, and by UPV-EHU (00039.310-15968/2004).
AR acknowledges support by the Joint Optical Metrology Center (JOMC),
Braunschweig. 
\end{acknowledgments}
%

\begin{appendix}

\section{Effective state-selective mirror potentials in 3D}

In this section we will generalize the results of \cite{ruschhaupt_2004_eeqt}
to the
three-dimensional case. The objective is to justify the dependence 
assumed in the main text for the state-selective mirror potentials.
We assume two levels connected with a detuned laser characterized
by a wave vector $k_L$ in $y$ direction, a Rabi frequency $\Omega (x)$,
and a large detuning $\Delta$.
We shall also assume as before that the atom impinges on the laser region in the 
ground state with wave number $k>0$.     
The wave vector $\Psi=\left(\psi_1\atop\psi_2\right)$
of the atom in the laser adapted interaction picture should
obey the stationary Schr\"odinger equation, 
\begin{eqnarray*}
{H}{\Psi} (x,y)=E{\Psi}(x,y)
\end{eqnarray*}
where now 
\begin{eqnarray*}
{H}=\frac{{p}^{\,2}}{2m} +
\frac{\hbar}{2}\left(
\begin{array}{cc}
0 & \Omega(x) e^{-i k_L y}\\
\Omega(x) e^{i k_L y} & -2\Delta
\end{array}
\right),
\end{eqnarray*}
$E=\hbar^2 k^2/2m$, ${p}^2={p}_x^2 + {p}_y^2$,
$\Omega(x)$ is the position dependent Rabi frequency,
$\Delta=\omega_{L}-\omega_{21}$
is the detuning between the laser frequency and the atomic transition 
frequency $\omega_{21}$.
In components, we have:
\begin{eqnarray}
E\psi_1(x,y) &=& \frac{{p}^{\,2}}{2m}\psi_1(x,y)\nonumber\\
& & \nonumber\\[-0.5cm]
& & + \frac{\hbar}{2}\Omega(x) e^{-i k_L y}\psi_2(x,y)\label{ham_2l_1}\\
E\psi_2(x,y) &=& \frac{{p}^{\,2}}{2m}\psi_2(x,y)
+ \frac{\hbar}{2}\Omega(x) e^{i k_L y}\psi_1(x,y)\nonumber\\
& &- \hbar \Delta \, \psi_2(x,y)\, .\label{ham_2l_2}
\end{eqnarray}
We can write Eq. (\ref{ham_2l_2}) in the following way
\begin{eqnarray*}
\left[E + \hbar\Delta 
- \frac{{p}^{\,2}}{2m}\right]\psi_2(x,y)
= \frac{\hbar}{2} \Omega(x) e^{i k_L y}\psi_1(x,y),
\end{eqnarray*}
and if $\Delta$ is large it follows heuristically that 
\begin{eqnarray*}
\psi_2(x,y) &\approx&
\frac{\Omega(x)}{2\Delta} e^{i k_L y}\psi_1(x,y).
\end{eqnarray*}
Putting this into equation (\ref{ham_2l_1}) we get 
\begin{eqnarray*}
E\psi_1(x,y)
& \approx & \frac{{p}^{\,2}}{2m} \psi_1(x,y)+
\frac{\hbar}{2} W (x) \psi_1(x,y),
\end{eqnarray*}
where
\begin{eqnarray*}
W (x)=\frac{\Omega(x)^2}{2\Delta},
\end{eqnarray*}
i.e., there results a local and energy-independent approximation
$W (x)$ to the exact optical potential for state 1. 
Note that $W (x)$ equals the one-dimensional effective potential derived
in \cite{ruschhaupt_2004_eeqt} and does only depend on $x$ even
in a three-dimensional case.
It is also remarkable that it is energy and angle independent, 
and these properties 
rely on the possibility to neglect kinetic energies and Doppler 
terms with respect to the high detuning.

\end{appendix}


\begin{thebibliography}{1}
\bibitem{rev} D. E. Pritchard, A. D. Cronin, S. Gupta, D. A. Kokorowski, { Ann. Phys. (Leipzig)} {\bf 10}, 35 (2001).

\bibitem{ruschhaupt_2004_diode}
A. Ruschhaupt and J. G. Muga, {\rm Phys. Rev.} A {\bf 70}, 061604(R) (2004).

\bibitem{ruschhaupt_2006_diode}
A. Ruschhaupt and J. G. Muga, { Phys. Rev. A} {\bf 73} (2006) 013608.

\bibitem{ruschhaupt_2006_quench} 
A. Ruschhaupt, J. G. Muga, and M. G. Raizen,
{J. Phys.} B: At. Mol. Opt. Phys. {\bf 39}, L133 (2006).

\bibitem{raizen.2005}
M. G. Raizen, A. M. Dudarev, Qian Niu, and N. J. Fisch,
{Phys. Rev. Lett.} {\bf 94}, 053003 (2005).

\bibitem{dudarev.2005}
A. M. Dudarev, M. Marder, Qian Niu, N. J. Fisch, and M. G. Raizen,
{Europhysics Letters} {\bf 70}, 761 (2005).

\bibitem{maxwell} J. C. Maxwell, {\it Theory of Heat, 4th edition} 
(London: Longmans, Green and Co., 1875), p. 328-329.
 
\bibitem{leff1}
H. S. Leff and A. Rex (eds), {\it Maxwell's Demon: entropy, information,
computing}
(Princeton: Princeton University Press, 1990).

\bibitem{leff2}
H. S. Leff and A. Rex (eds),
{\it Maxwell's Demon 2: entropy, classical and quantum information,
computing} (Bristol and Philadelphia: Institute of Physics Publishing, 2003).

\bibitem{ruschhaupt_2006_demon}
A. Ruschhaupt, J. G. Muga, and M. G. Raizen,
{\it J. Phys. B} {\bf 39}, 3833 (2006).

\bibitem{bergmann.1998}
K. Bergmann, H. Theuer, and B. W. Shore,
{\rm Rev. Mod. Phys.} {\bf 70}, 1003 (1998).

\bibitem{ruschhaupt_2004_eeqt}
A. Ruschhaupt, J. A. Damborenea, B. Navarro, J. G. Muga, and
G. C. Hegerfeldt, {Europhys. Lett.} {\bf 67}, 1 (2004).

\bibitem{qj}  G. C. Hegerfeldt and T. S. Wilser in:
{\it Classical and Quantum Systems.} 
Proceedings of the Second International Wigner Symposium, July
1991, edited by H.~D. Doebner, W. Scherer, and F. Schroeck, (Singapore: World
Scientific, 1992), p. 104;
%
G. C. Hegerfeldt,
{\it Phys. Rev. A} {\bf 47} 449 (1993);
%
G. C. Hegerfeldt and D. G. Sondermann, {\it Quantum 
  Semiclass.~Opt.} {\bf 8} 121 (1996).
%
For a review cf.
M. B. Plenio and P. L. Knight
{\it Rev. Mod. Phys.} {\bf 70} 101 (1998).
The quantum jump approach is
essentially equivalent to the Monte-Carlo wave function approach of 
J. Dalibard, Y. Castin, and  K. M{\o}lmer K 1992  
    {\it Phys. Rev. Lett.} {\bf 68} 580 (1992),
%
and to the quantum trajectories of 
H. Carmichael, {\em An Open Systems Approach to Quantum 
Optics}, Lecture Notes in Physics m18 (Berlin: Springer, 1993).


\end{thebibliography}
\end{document}